\documentclass[11pt]{article}
\usepackage{geometry}                
\usepackage{graphicx}
\usepackage{lmodern}
\usepackage{revsymb}
\usepackage{amssymb}
\usepackage{mathdots}
\usepackage{epstopdf}
\usepackage{setspace}
\usepackage{amsmath}
\usepackage{amsthm}
\usepackage{fancyref}
\usepackage{fancyhdr}
\usepackage{authblk}
\usepackage{cite}

\usepackage{tikz}
\usepackage{float}

\newcommand{\C}{\mathbb{C}}                   
\renewcommand{\bar}[1]{\overline{#1}}         

\newtheorem{theorem}{Theorem}

\newtheorem{corollary}[theorem]{Corollary}
\newtheorem{proposition}{Proposition}
\newtheorem{definition}{Definition}

\newtheorem{lemma}{Lemma}

\newtheorem{remark}{Remark}

\newcommand{\X}{\mathcal{X}}
\newcommand{\Y}{\mathcal{Y}}
\newcommand{\Z}{\mathcal{Z}}
\newcommand{\W}{\mathcal{W}}

\newcommand{\Choi}{\mathcal{C}}

\newcommand\tr{\mathop{\rm Tr}\nolimits}

\title{A Characterization of Antidegradable Qubit Channels}

\author[1,2]{Connor Paddock}
\author[3]{Jianxin Chen}

\affil[1]{\emph{Department of Combinatorics \& Optimization, University of Waterloo}}
\affil[2]{\emph{Institute for Quantum Computing, University of Waterloo}}
\affil[3]{\emph{Aliyun Quantum Laboratory}}

\date{}                                           

\begin{document}

\maketitle

\begin{abstract}
This paper provides a characterization for the set of antidegradable qubit channels. The characterization arises from the correspondence between the antidegradability of a channel and the symmetric extendibility of its Choi operator. Using an inequality derived to describe the set of bipartite qubit states which admit symmetric extension, we are able to characterize the set of all antidegradable qubit channels. Using the characterization we investigate the antidegradability of unital qubit channels and arbitrary qubit channels with respect to the dimension of the environment. We additionally provide a condition which describes qubit channels which are simultaneously degradable and antidegradable along with a classification of self-complementary qubit channels.
\end{abstract}

\section{Introduction}

This paper is concerned with the characterization of \emph{degradable} and \emph{antidegradable single-qubit channels}. Degradable channels were introduced as a class of channels which have additive coherent information \cite{devetak2005capacity}, a feature which allows for the \emph{quantum channel capacity} to be explicitly calculated. The notion of an antidegradable channel comes from the converse definition of a degradable channel. Degradable qubit channels were first studied in the context of \emph{small environments} in \cite{wolf2007quantum}. The authors found that the degradability and antidegradability of a channel is highly dependent on the dimension of the environment. Following \cite{wolf2007quantum}, the mathematical structure of degradable and antidegradable maps was extensively studied \cite{cubitt2008structure} using the qubit channel Bloch sphere parametrization of \cite{king2001minimal,ruskai2002analysis}. However, a characterization for the set of non-unital antidegradable channels remained to be described.

To develop a full characterization of degradable and antidegradable qubit channels we use the correspondence between antidegradable channels and symmetric extensions of the qubit channel Choi operator \cite{myhr2009spectrum}. Specifically, we use the derivation on an inequality from \cite{chen2014symmetric} which gives a necessary and  sufficient condition for symmetric extendibility of bipartite qubit states. Via the connections between states and channels, the inequality can be used to completely characterize all antidegradable qubit channels.\\
\begin{corollary}\label{anti_cor}
A singe qubit channel $\Phi$ is antidegradable if and only if the following inequality holds for a positive semi-definite Choi matrix $\Choi_\Phi$,
\begin{equation}\label{eq:anti_cor}
\det \big(\Choi_\Phi\big)\geq \bigg(\frac{\tr\left(\Choi_\Phi^2\right)+\tr \big(\Phi(I)^2\big)}{4}\bigg)^2 ,
\end{equation}
where $I$ is the identity matrix on $\C^2$.
\end{corollary}

We demonstrate how this new characterization can be used an alternative method to obtain many results from \cite{wolf2007quantum,Leung:2015aa} and \cite{cubitt2008structure}. We also establish some remarks for antidegradable qubit channels with various dimensions of the environment system, and some properties for self-complementary channels. This result along with the appropriate complete positivity condition for qubit channels provides easily \emph{computable} conditions for the antidegradbility of the remaining case of non-unital qubit channels. We briefly discuss some results for the set of self-complementary and simultaneously degradable/antidegradable qubit channels, in addition to some results on the antidegradability of the qubit dephasing, qubit amplitude damping, and qubit depolarizing channel.

Unless stated otherwise we will assume that we are in a finite dimensional inner product space over the field of complex numbers, which we will denote $\C^n$. We also assume we are working with the algebra of bounded, linear operators on $\C^n$, denoted by $\mathcal{L}(\C^n)$, where each linear operator has a representation in the set of complex valued matrices $\mathcal{M}(\C^n)$. As above we will use the symbol $I$ to denote the $n \times n$ identity matrix on $\C^n$, while reserving $\mathcal{I}$ to denote the identity map on the set of bounded linear operators $\mathcal{L}(\C^n)$. We will reserve script letters $\W,\X,\Y,\Z $, to be  Euclidean spaces $\C^n$. While $\mathcal{U},\mathcal{V},Psd,\mathcal{D}$, will be reserved to represent the space of unitaries, isometries, positive semi-definite, and density matrices respectively. A number of useful representations in quantum information make use of the \emph{vectorization} of an operator. The vectorization or $vec$, is a bijective map between operators, and vectors in the tensor product of the output space, with the input space of the operator. Unfortunately, there is more than one convention for this map, so we will be following that used in \cite{horn1991topics}, whereby one can think of $vec(A)$ as the mapping which stacks the columns of the $n \times m$ matrix $A$ on top of each other, to obtain a single column vector of dimension $nm$.

\subsection{Symmetric Extensions, and k-Extendibility}

From here on we will use $\rho_{\X\Y}$ to denote a state $\rho \in \mathcal{D}(\X \otimes \Y)$. Recall that for any given $\rho_{\X\Y}$, the states represented by $\tr_\X(\rho_{\X\Y})=\rho_\Y $ and $\tr_\Y(\rho_{\X\Y})=\rho_\X $ are called the marginal states of $\rho_{\X\Y}$.
One can consider the marginals of a state supported on an arbitrary number of product product spaces. For example a state $\sigma \in \mathcal{D}(\X^{\otimes{n}})$ will admit a marginal states, each one from the appropriate tracing out the other $1\leq k \leq n-1$ spaces. Marginal of a quantum state will become fundamental in the following context of symmetric extensions.

\begin{definition}
A bipartite state $\rho_{\X\Y}$ is said to admit a symmetric extension if there exists a space $\Y'$ such that the tripartite state $\rho_{\X\Y\Y'}$ admits the same marginal states,
\begin{equation}\label{eq:sym_ext}
\tr_\Y(\rho_{\X\Y\Y'})=\rho_{\X\Y'}=\rho_{\X\Y}=\tr_{\Y'}(\rho_{\X\Y\Y'})\;.
\end{equation}
\end{definition}

For qubits the following theorem characterizes when a density operator admits a pure state symmetric extension through its spectrum denoted $spec(\rho)$ \cite{myhr2009spectrum}.

\begin{lemma}[Myhr \& Lutkenhaus \cite{myhr2009spectrum}]
Let $\rho \in \mathcal{D}(\X \otimes \Y)$ be a bipartite two qubit state, $dim(\X)=dim(\Y)=2$, then $\rho$ admits a pure state symmetric extension if and only if $\rho$ is an element of the following set
\begin{equation}\label{eq:sym_ext_lem}
\mathcal{A}=\left\{ \rho_{\X\Y} : spec(\rho_{\X\Y})=spec(\rho_\Y) \right \} \, .
\end{equation}
\end{lemma}

It was later shown that we can characterize not only the states which admit pure extension, but any symmetric extension.

\begin{theorem}[Myhr \& Lutkenhaus \cite{myhr2009spectrum}]\label{bipartite_symext}
Let $\rho_{\X\Y}$ be a bipartite qubit state then $\rho$ admits a symmetric extension if and only if $\rho$ is an element of the following set
\begin{equation}\label{eq:convA}
conv(\mathcal{A})=conv\left\{ \rho_{\X\Y} : spec(\rho_{\X\Y})=spec(\rho_\Y) \right \} \, ,
\end{equation}
where $conv(\mathcal{A})$ is the convex hull of $\mathcal{A}$.
\end{theorem}


\begin{definition}
A state $\rho_{\X\Y}$ admits a $k$-symmetric extensions $\rho_{\X\Y\Y^{(1)}\ldots{\Y^{(t)}}\ldots{\Y^{(k)}}}$ if the marginal state obtained by tracing out all but any one $\Y$ space $\Y^{(t)}$ for $1 \leq t\leq k$ is equivalent to the state $\rho_{\X\Y}$.
\end{definition}

If a state is $k$-symmetric extendible for all $k \in \mathbb{N}$ we say the state is \emph{exhaustively} extendible. An important note is that a state is $k$-symmetric extendible for all $k\in\mathbb{N}$ if and only if it is separable \cite{fannes1988symmetric,werner1989application}.

\subsection{Complementary Quantum Channels}

In the study of quantum information a quantum channel models the way in which two parties send and receive quantum information. Mathematically a quantum channel is a completely positive trace preserving $(CPTP)$ linear map $\Phi: \mathcal{L}(\X)\mapsto \mathcal{L}(\Y)$. We will denote the set of quantum channels $\Phi \in C(\X,\Y)$, for which we mean $\Phi$ is the $CPTP$ map from $\mathcal{D}(\X)$ to $\mathcal{D}(\Y)$. The representations are due primarily to the work of Stinespring \cite{stinespring1955positive}, Choi \cite{choi1975completely}, Jamio{\l}kowski \cite{jamiolkowski1972linear}, and Kraus.

Recall that the Stinespring representation of a channel has an auxiliary space which induces what is known as the \emph{complementary} channel.

\begin{definition}\label{comp}
Two channels $\Psi\in C(\X,\Y)$ and $\widetilde{\Psi}\in C(\X,\Z)$ are said to be complementary if there exists an isometry $V\in \mathcal{V}(\X,\Y\otimes \Z)$, such that the Stinespring representations of the two channels are of the form
\begin{equation}\label{eq:comp}
\Psi(\rho)=\tr_\Z(V\rho V^*)\text{ and }\widetilde{\Psi}(\rho)=\tr_\Y(V\rho V^*)\; .
\end{equation}
\end{definition}

We denote $\widetilde{\Phi}$ the complementary channel of $\Phi$. Furthermore, it was shown in \cite{holevo2007complementary,king2005properties} that if a channel has the Kraus representation then the following proposition holds.

\begin{remark}[Holevo, et. al \cite{holevo2007complementary,king2005properties}]
If a quantum channel has a Kraus representation with a set of operators $\{K_i\}_{i=1}^d$ then the complementary channel is
\begin{equation}\label{eq:kraus_complement}
\widetilde{\Phi}(\rho)=\sum_{i,j}^d  \tr(\rho K_j^*K_i)E_{i,j}\; .
\end{equation}
\end{remark}

Recall that the Choi matrix is positive semi-definite and hence we can consider a purifications of the Choi matrix. In fact the Choi matrix of both a channel and its complementary channel are purified by the same vector $vec(V) \in (\X\otimes\Y\otimes \Z)$.

\subsection {Qubit Channels and Bloch Sphere Transformations}
We call a channel a qubit channels, by which we mean a \emph{single-qubit} channel which maps some state in $\mathcal{D}(\X)$ into $\mathcal{D}(\Y)$, where $dim(\X)=dim(\Y)=2$.

Because the qubit is only a two dimensional object the representations admitted are fairly easy to work with. We recall that from earlier that every qubit state is isomorphic to a vector in the unit ball of $\mathbb{R}^3$. Every linear, completely positive, trace preserving map $\Phi$ on the space of $2\times 2$ density matrices $\rho$ can be represented in the form,

\begin{equation}\label{eq:qubit_channel}
\Phi(\rho)=U\Lambda(V{\rho}V^*)U^*\;,
\end{equation}

where $U$ and $V$ are unitary operators, and $\Lambda$ is a qubit channel with the \emph{natural} representation in the Pauli basis. To see how we obtain the unitary $U$, we first consider the action of a qubit channel on the Bloch sphere representation of a qubit state. In that case the channel is given by some linear operator $\mathcal{T}$ such that

\begin{equation}
\Lambda(\rho)\mapsto \mathcal{T}(\tfrac{1}{2}[I+w\cdot \sigma])=\tfrac{1}{2}(I+(t+Tw)\cdot \sigma)
\end{equation}

in other words we can decompose $\mathcal{T}$ into the block operator $\mathcal{T}=\begin{pmatrix}
1 & 0\\
t & T \end{pmatrix}\;,$

where $t \in \mathbb{R}^3$ and $T \in \mathcal{M}(\mathbb{R}^3)$ is a $3 \times 3$ matrix. The vector $t$ determines the translation of the map about the identity, which is the origin of the Bloch ball, this implies that a qubit channels is unital if and only if $t=0$. Diagonalizing the matrix $T$ leads to the form of $\mathcal{T}_\Lambda$ in the basis of the Pauli matrices as,

\begin{equation}\label{eq:qubit_matrix}
\mathcal{T}_{\Lambda}=\begin{pmatrix}
1 & 0 & 0 & 0 \\
t_1 & \lambda_1 & 0 & 0 \\
t_2 & 0 & \lambda_2 & 0 \\
t_3 & 0 & 0&  \lambda_3 \\
\end{pmatrix} \;.
\end{equation}

Complete positivity for the unital map is given by the following necessary condition on the values of $T$. Each $\lambda_i$ for $i=1,2,3$ must be in the tetrahedron $\mathbf{T}$, given by
\begin{equation}\label{eq:tetrahedron}
\mathbf{T}=conv\left\{(1,1,1),(1,-1,-1),(-1,1,-1),(-1,-1,1)\right\} \;,
\end{equation}
this fact is enough to ensure complete positivity of the map in the unital case. In the non-unital case, complete positivity of the map is given by the more complex Algoet-Fujiwara conditions \cite{fujiwara1999one}. Hence, every qubit channel unitarily equivalent to a qubit channel $\Lambda$ defined by the parameters in the above matrix representation.

In fact one can obtain a parametrized Choi-Jamio{\l}kowski representations for all qubit channels $\Lambda$ up to isomorphism,

\begin{equation}\label{eq:qubit_choi}
\Choi_\Lambda=\frac{1}{2}\begin{pmatrix}
1+t_3+\lambda_3 & t_1-it_2 & 0 & \lambda_1+\lambda_2 \\
t_1+it_2 & 1-t_3-\lambda_3 & \lambda_1-\lambda_2 & 0 \\
0 & \lambda_1-\lambda_2 & 1+t_3-\lambda_3 & t_1-it_2 \\
\lambda_1+\lambda_2 & 0 & t_1+it_2 & 1-t_3+\lambda_3
\end{pmatrix}
\end{equation}

We note that \eqref{eq:qubit_choi} was first calculated in \cite{ruskai2002analysis}. We remark that the Choi matrix of a qubit channel has a slightly nicer form form in the Bell basis, obtained by applying the unitary transform $F\Choi_\Lambda F^*$ \cite{braun2014universal}.

\begin{equation}\label{eq:bell_choi}
F=\frac{1}{\sqrt{2}} \begin{pmatrix}
1 & 0 & 0 & 1 \\
0 & 1 & 1 & 0 \\
0 & 1 & -1 & 0  \\
1 & 0 & 0 & -1
\end{pmatrix}\, \text{hence},\; F\Choi_\Lambda F^*=\frac{1}{2}\begin{pmatrix}
\mu_0 & t_1 & -it_2 & t_3 \\
t_1 & \mu_1 & -t_3 & it_2 \\
it_2 & -t_3 & \mu_2 & t_1 \\
t_3 & -it_2 & t_1 & \mu_3
\end{pmatrix}
\end{equation}

where, $\mu_0=(1+\lambda_1+\lambda_2+\lambda_3)$, $\mu_1=(1+\lambda_1-\lambda_2-\lambda_3)$,  $\mu_2=(1-\lambda_1+\lambda_2-\lambda_3)$, $\mu_3=(1-\lambda_1-\lambda_2+\lambda_3) $.\\

In the case of the unital qubit channel are merely a restriction of the general case to when $t_i=0$ for $i=1,2,3$, abbreviated we write a unital qubit channel as $\mathbb{U}=\Lambda|_{t=0}$, hence

\begin{equation}\label{eq:unital_qubit}
\mathcal{T}_{\mathbb{U}}=\begin{pmatrix}
1 & 0 & 0 & 0 \\
0 & \lambda_1 & 0 & 0 \\
0 & 0 & \lambda_2 & 0 \\
0 & 0 & 0&  \lambda_3 \\
\end{pmatrix}\;,
\end{equation}

which reduces the number of parameters to three in the Choi matrix. Furthermore, the Choi matrix for the unital qubit channels is of the form,

\begin{equation}\label{eq:unital_choi}
\Choi_\mathbb{U}=\frac{1}{2}\begin{pmatrix}
1+\lambda_3 & 0 & 0 & \lambda_1-\lambda_2 \\
0 & 1-\lambda_3 & \lambda_1+\lambda_2 & 0 \\
0 & \lambda_1+\lambda_2 & 1-\lambda_3 & 0 \\
\lambda_1-\lambda_2 & 0 & 0 & 1+\lambda_3
\end{pmatrix}\;.
\end{equation}

Applying the unitary transform such that the matrix $\Choi_\mathbb{U}$ is in the Bell basis one obtains the diagonal Choi matrix for unital qubit channels,

\begin{equation}\label{eq:bell_unital_choi}
F\Choi_\mathbb{U}F^*=\frac{1}{2}\begin{pmatrix}
\mu_0 & 0 & 0 & 0 \\
0 & \mu_1 & 0 & 0 \\
0 & 0 & \mu_2 & 0 \\
0 & 0 & 0 & \mu_3
\end{pmatrix}\;,
\end{equation}

where each $\mu_i$ is the same as in \eqref{eq:bell_choi}.\\

\subsection{Degradable and Antidegradable Quantum Channels}

Recalling the definition of the complementary channel \eqref{eq:comp}, we now present the definition of degradable and antidegradable channels.

\begin{definition}
A channel $\Phi$ is said to be degradable if there exits a channel $\Gamma$, such that the complementary channel $\widetilde{\Phi}$ can be emulated by a composition of $\Gamma$  with the original channel $\Phi$,
\begin{equation}
\widetilde{\Phi}=\Gamma \circ \Phi.
\end{equation}
\end{definition}

\begin{figure}[H]
\begin{center}
 \begin{tikzpicture}[scale=0.8]
\path (0,0) node(x) {\large$\mathcal{X}$} 
	(3,-2) node(y) {\large$\mathcal{Y}$}
	(6,2) node(w)  {\large$\mathcal{W}$}
      (3,2) node(z) {\large$\mathcal{Z}$};
\draw[->] (x) -- (y)
node[pos=0.5,below] {\Large$\Phi$};
\draw[->] (x) -- (z)
node[pos=0.5,above] {\Large$\widetilde{\Phi}$};
\draw[->] (y) -- (z)
node[pos=0.5,right] {\Large$\Gamma$};
\draw[->] (y) -- (w);
\end{tikzpicture}
\end{center}
\caption{Diagram of a degradable channel $\Phi \in C(\X,\Y)$ with auxiliary space $\Z$, where $\Z \otimes \mathcal{W}$ is the image space of the isometry induced by the degrading channel $\Gamma$. \cite{holevo2012book}}
\end{figure}
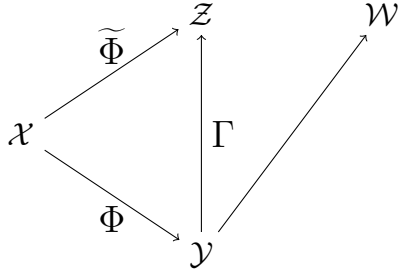

The notion of a degradable channel is that there is some other channel which adds a particular type of \emph{noise} to the channel such that it can emulate the complementary channel. The converse of a degradable channel is an \emph{antidegradable} channel. In that case there exists a particular channel which can add noise to the complementary channel such that it emulates the original channel.

\begin{definition}
A channel is said to be antidegradable if there exists a channel $\Xi$, such that $\Psi$ can be emulated by a composition of $\Xi$ and the complementary channel $\widetilde{\Psi}$,
\begin{equation}
\Psi= \Xi \circ \widetilde{\Psi}
\end{equation}
\end{definition}

\begin{proposition}\label{degrad_antidegrad}
A channel is degradable if and only if its complementary channel is antidegradable.
\end{proposition}

Degradable channels first appeared in \cite{devetak2005capacity}, as channels with strictly additive \emph{coherent information}, which leads to a simple mathematical expression of the quantum capacity. Many of these facts for degradable channels were first noted in \cite{cubitt2008structure,devetak2005capacity}. A large set of channels which are degradable are the \emph{dephasing} channels, which were also defined in \cite{devetak2005capacity}. The idea of a dephasing channel was most likely understood prior to this as a channel which destroys the quantum states \emph{phase}, making them more classical. Here we choose our preferred basis to be the standard basis $\{e_i\}_{i=1}^n$.

\begin{definition}\label{dephasing}
A channel $\Psi \in C(\X,\Y)$ for which a preferred orthonormal basis $\{e_i\}_{i=1}^n \in \X$ is preserved by the partial isometry of the Stinepring representation,
\begin{equation}\label{eq:dephasing}
U:e_i\mapsto e_i \otimes u_i\;,
\end{equation}
where $\{u_i\}_{i=1}^n$ is a set of not necessarily orthogonal vectors in $\Z$, is said to be a dephasing channel.
\end{definition}

\begin{definition}
When in addition to being a dephasing channel, if the basis of the auxiliary space $\{u_i\}_{i=1}^n \in \Z$ consists of mutually orthogonal vectors we obtain the completely dephasing channel, denoted by $\Delta$, such that,
\begin{equation}\label{eq:comp_dephase}
\Delta(\rho)=\sum_{i=1}^n \rho_{i,i} E_{i,i}\;,
\end{equation}
where $\rho_{i,i}$ are the diagonal entries of the matrix $\rho$.
\end{definition}

\begin{lemma}[Devetak \& Shor \cite{devetak2005capacity}]\label{dephase_props}
If a channel is dephasing, then it satisfies the following properties:
\begin{enumerate}
\item $\Psi \circ \Delta=\Delta \circ \Psi=\Delta$
\item $\widetilde{\Psi} \circ \Delta=\widetilde{\Psi}$
\end{enumerate}
\end{lemma}

\textbf{Proof:}
The first property is proved by the fact that the completely dephasing channel is the map which takes all off-diagonal entries to zero and leaves the diagonal entries untouched. Where as in general a dephasing channel may only partially take the off-diagonal entries to zero, and leave the diagonal entries unaltered. For the second property we remark that the complementary of a dephasing channel admits the following Stinespring representation through the structure of the partial isometry in \eqref{eq:dephasing},
\begin{equation}\label{eq:dephase_comp}
\begin{aligned}
\widetilde{\Psi}(\rho)=& \tr_\Y(U\rho U^*) \\ =&\tr_\Y\bigg(\left(\sum_i^n (e_i \otimes u_i)e_i^*\right) \rho \left(\sum_j^n (e_j \otimes u_j)e_j^*\right)^* \bigg)\\
=&\tr_\Y\left(\sum_{i,j}^n (e_i \otimes u_i)e_i^* \rho e_j(e_j^* \otimes u_j^*) \right)\\
=&\tr_\Y\left(\sum_{i,j}^n \langle e_i, \rho e_j \rangle (e_i e_j^* \otimes u_iu_j^*) \right)\\ =&\sum_{i,j}^n \langle e_i, \rho e_j \rangle \tr(e_i e_j^*) u_iu_j^* \\
=&\sum_{i}^n \langle e_i, \rho e_i \rangle u_iu_i^*.
\end{aligned}
\end{equation}
We notice the action of the complementary channel is only dependent on the diagonal entries of the state $\rho$ and hence, first applying the completely dephasing channel does not affect the output of the complementary channel. $_\blacksquare$\\

From the above it follows that every dephasing channel is degradable. Furthermore, dephasing channels are degraded by their own complementary channel.

\begin{equation}
\widetilde{\Psi}=\widetilde{\Psi} \circ \Delta=\widetilde{\Psi} \circ \Delta \circ \Psi=\widetilde{\Psi} \circ \Psi.
\end{equation}

Remark that if we chose to trace out $\Z$ in \eqref{eq:dephase_comp}, instead of $\Y$, we obtain a general expression for a dephasing channel, 
\begin{equation}\label{eq:dephase_channel}
\sum_{i,j}^n \langle e_i, \rho e_j \rangle \tr(u_iu_j^*)E_{i,j}
\end{equation}
with respect to the basis $\{e_i\}_{i=1}^n$.

Another property of dephasing channels is that when the dimension of the complementary space is equal to that of the input space, the set of Kraus operators in their operator-sum representations are simultaneously diagonalizable. This is why in the past they have been called ``twisted-diagonal'', or ``diagonal'' maps in some literature. Like the dephasing channels for degradability an important set of antidegradable quantum channels are known as \emph{entanglement breaking} channels. 

\begin{definition}\label{ebt}
A channel $\Phi \in C(\X,\Y)$ is entanglement breaking if for any state $\rho \in \mathcal{D}(\W\otimes\X)$ the output $(\mathcal{I} \otimes \Psi)(\rho) \in \mathcal{D}(\W \otimes \X)$ is separable.
\end{definition}

\begin{proposition}
Every entanglement breaking channel is antidegradable.
\end{proposition}

\textbf{Proof:}
We first remark that for a channel to be entanglement breaking it must have full Choi rank, otherwise the channel will not break the entanglement of the state $dim(\X)^{-1}(vec(I)vec(I)^*)$ \cite{horodecki2003entanglement}.
The proof follows from proposition \ref{degrad_antidegrad}, and the fact that every entanglement breaking channel can be written in terms of rank 1 Kraus operators, then the complementary channel must have diagonal Kraus operators, hence it is a dephasing map, which is degradable.

More precisely, due to the form of the entanglement breaking channels and the fact that we can obtain the complementary channel from the fact that entanglement breaking channels have the form \eqref{eq:kraus_complement} one can see that,
\begin{equation}
\begin{aligned}
\widetilde{\Phi}(\rho)=&\sum_{i,j}^d \tr(u_iv_i^*\rho v_ju_j^*)e_ie_j^*\\=&\sum_{i,j}^d  \langle{v_i,\rho v_j}\rangle \tr({u_iu_j^*}) e_ie_j^*\\=&\sum_{i,j}^d  \langle{v_i,\rho v_j}\rangle \tr({u_iu_j^*}) E_{i,j}\;.
\end{aligned}
\end{equation}
We note that if one chooses the basis $\{v_i\}_{i=1}^d=\{e_i\}_{i=1}^d$, then we obtain a channel of the form \eqref{eq:dephase_channel}, hence we see the complementary channel is a dephasing map which are degradable, hence by proposition \ref{degrad_antidegrad} the result follows. $_\blacksquare$\\

\section{Antidegradable Channels and Symmetric Extensions of the Choi Operator}

Most cases for qubit channel degradability and antidegradability were fully characterized in \cite{wolf2007quantum,cubitt2008structure}. The non-unital antidegradable qubit channels remained to be characterized. We take the suggestion of \cite{chen2014symmetric} and \cite{myhr2009spectrum} and show that the spectrum criteria of symmetrically extendable bipartite qubit states provides a characterization for all antidegradable qubit channels. It was previously noted that the Choi-Jamio{\l}kowski representation can be used to characterize a large class of antidegradable channels namely the set of entanglement breaking channels, by separability of the Choi operator \cite{horodecki2003entanglement}. When we recall that symmetric extendability is a \emph{weak} notion of separability, it is not surprising that we apply symmetric extendibility of the Choi matrix to characterizing antidegradability. This idea was made explicit by the following lemma in \cite{myhr2009spectrum}.\\

\begin{lemma}[Myhr \& Lutkenhaus \cite{myhr2009spectrum}]\label{anti_sym}
A channel $\Psi$ is antidegradable if and only if its Choi-Jamio{\l}kowski representation admits a symmetric extension.
\end{lemma}

Recall that from the definition \ref{ebt} every entanglement breaking channel has a separable Choi matrix. Recall that any separable operator will admit a symmetric extension, in fact we recall that it will be \emph{exhaustively} symmetrically extendable. Hence, lemma \ref{anti_sym} provides an alternate proof that entanglement breaking channels are antidegradable.

\subsection{Analytic Condition for Bipartite Qubit State Symmetric Extendability}

The connection between the antidegradability of a qubit channel and bipartite qubit state symmetric extendibility, is provided by the fact that Choi operators for qubit channels are in one-to-one correspondence with bipartite qubit states through the Choi-Jamio{\l}kowski isomorphism. The question of whether a bipartite state admits a symmetric extension is non-trivial. A criteria for the spectrum of the state and its marginal was conjectured in \cite{myhr2009spectrum} to describe the set of symmetrically extendible bipartite qubit states \eqref{eq:convA}. The equivalence of states which satisfy \eqref{eq:sym_ext_thm} and states in $\mathcal{A}$ \eqref{eq:convA} was conjectured by noticing that their condition on the spectrum held for every case they investigated \cite{myhr2009spectrum}. Shortly after the conjecture was proved in \cite{chen2014symmetric}.

\begin{theorem}[Chen, et. al \cite{chen2014symmetric}]\label{sym_ext_thm}
A two qubit state $\rho_{AB}$ admits a symmetric extension if and only if,
\begin{equation}\label{eq:sym_ext_thm}
\tr(\rho^2_B)\geq \tr(\rho^2_{AB})-4\sqrt{\det(\rho_{AB})}
\end{equation}
\end{theorem}

The inequality \eqref{eq:sym_ext_thm} can also be regarded as an inequality involving the arithmetic mean of the square of the eigenvalues for the marginal state $\rho_B$, and the difference between the arithmetic and geometric mean of the square eigenvalues for the full state $\rho_{AB}$. We note that the equivalence provided by lemma \ref{anti_sym} implies that the set of antidegradable channels is convex, due to convexity of the set of  symmetrically extendible states. Convexity of the set of antidegradable channels was alternatively demonstrated in \cite{cubitt2008structure}. Interestingly, the authors of \cite{cubitt2008structure} also showed that the set of degradable channels is not convex. 

We return to theorem \ref{sym_ext_thm}, which we now give as a statement about the Choi operator using lemma \ref{anti_sym}.

\begin{corollary}\label{anti_cor}
A qubit channel $\Phi$ is antidegradable if and only if the following inequality holds for the Choi operator of a single-qubit channel $\Choi_\Phi$.
\begin{equation}\label{eq:anti_cor}
\tr \big(  \Phi(I)^2\big) \geq \tr\left(\Choi_\Phi^2\right)-4\sqrt{\det \big(\Choi_\Phi\big)},
\end{equation}
and the Choi operator is positive semidefinite.
\end{corollary}

This is obtained from the remark that $\tr_{\X}( \Choi_\Phi)=\Phi(I)$. The implication of corollary \ref{anti_cor} is that the set of Choi operators which satisfy inequality \eqref{eq:anti_cor} is convex with respect to the single-qubit channels \cite{cubitt2008structure}.

We now consider the set of all qubit channels under the block sphere parametrization \eqref{eq:qubit_matrix}. First we establish sufficiency for the symmetric extendability criteria based on the values in \eqref{eq:qubit_choi} up to unitary conjugation.

\begin{lemma}\label{trace_lem}
Let $\mathcal{X}$ and $\mathcal{Y}$ be two finite dimensional complex Hilbert spaces, $U\in \mathcal{U}(\X)$, $V\in \mathcal{U}(\Y)$ and $M\in \mathcal{L}(\X \otimes \Y)$.  Then $\tr_\X((V^\top \otimes U)M((V^\top \otimes U)^*)=U\tr_\mathcal{X}(M)U^*$.
\end{lemma}

\textbf{Proof:} Consider the following decomposition, $M=\sum_i^r vec(A_i)vec(B_i)^*$, then using properties of the $vec$ map we can see that,
\begin{equation}
\begin{aligned}
&\tr_\X((V^\top \otimes U)M((V^\top \otimes U)^*)\\
=&\sum_i^r  \tr_\X((V^\top \otimes U)vec(A_i)vec(B_i)^*(V^\top \otimes U)^*)\\
=&\sum_i^r  \tr_\X(vec(UA_iV)vec(UB_iV)^*)\\
=&\sum_i^r  UA_iVV^*B_i^*U^*\\
=&U\left(\sum_i^r  A_iB_i^*\right)U^*\\
=&U\left(\tr_\X\big(\sum_i^r vec(A_i)vec(B_i)^*\big) \right)U^*\\
=&U\tr_\X(M)U^* \; . \;_\blacksquare
\end{aligned}
\end{equation}

Because the Choi representation \eqref{eq:qubit_choi} is in terms of the channel in the Pauli basis $\Lambda$ and we need the following proposition.

\begin{proposition}\label{unitary_reduction}
If $\Choi_\Phi$ has a symmetric extension then $\Choi_{U{\Lambda}U^*}$ has a symmetric extension.
\end{proposition}

\textbf{Proof:} If the channel $\Lambda$ has following Kraus representation $\Lambda(\rho)=\sum_i{K_i}{\rho}K_i^*$ then the channel $\Phi$ has the following Kraus representation,

\begin{equation}
\Phi(\rho)=\sum_iU{K_i}V{\rho}V^*K_i^*U^*\;.
\end{equation}

It follows that the Choi-Jamio{\l}kowski of $\Phi$ is,

\begin{equation}\label{eq:block_unitary}
\begin{aligned}
\Choi_\Phi&=\sum_ivec(U{K_i}V)vec(U{K_i}V)^*\\
&=\sum_i(V^\top \otimes U)vec({K_i})vec(K_i)^*(V^\top \otimes U)^*\\
&=(V^\top \otimes U)\Choi_\Lambda(V^\top \otimes U)^*\;. \quad _\blacksquare
\end{aligned}
\end{equation}

Because the transformation \eqref{eq:block_unitary} between the two Choi-Jamio{\l}kowski representations is a block unitary transform the eigenvalues are preserved.

\begin{corollary}
The Choi-Jamio{\l}kowski representation of a single-qubit channel $\Choi_\Lambda$ admits a symmetric extension if and only if,
\begin{equation}\label{eq:choi_sym_thm}
\tr \big(  \Lambda(I)^2\big) \geq \tr \big( \Choi_\Lambda^2 \big)-4\sqrt{\det(\Choi_\Lambda)}
\end{equation}
\end{corollary}

\textbf{Proof:}
This follows directly from lemma \ref{trace_lem} and proposition \ref{unitary_reduction} that the condition for symmetric extendibility of an arbitrary map, we only need to consider the Choi matrix \eqref{eq:qubit_choi} which arises from the map \eqref{eq:qubit_channel}.

\begin{corollary}
The single-qubit channel $\Lambda$ is antidegradable if and only if the channel satisfies \eqref{eq:choi_sym_thm}.
\end{corollary}

\subsection{Conditions for Antidegradable Unital Qubit Channels}\label{sec:unital}

We note that in the unital case the right hand side of the inequality \eqref{eq:choi_sym_thm} has an even simpler form.

\begin{corollary}
A unital qubit channel $\mathbb{U}:\mathcal{D}(\mathbb{C}^2)\mapsto \mathcal{D}(\mathbb{C}^2)$ is anti-degradable if its Choi operator is positive semidefinite and,
\begin{equation}
2\geq \sum_i \left(\mu_i^2\right)-4\left(\prod_i\mu_i^2\right)^{\frac{1}{4}}\,,\quad 0\leq i\leq3,
\end{equation}
where $\mu_0=(1+\lambda_1+\lambda_2+\lambda_3)$, $\mu_1=(1+\lambda_1-\lambda_2-\lambda_3)$,  $\mu_2=(1-\lambda_1+\lambda_2-\lambda_3)$, $\mu_3=(1-\lambda_1-\lambda_2+\lambda_3)$, each $\lambda_i$ is from the unital Choi representation \eqref{eq:unital_qubit}.
\end{corollary}

\textbf{Proof:}
Firstly we note that the left hand side of \eqref{eq:choi_sym_thm} reduces when our channel $\mathbb{U}$ is unital,
\begin{equation}
\tr \big( \mathbb{U}(I)^2\big)=\tr \big( I^2\big)=\tr \big( I\big)=2 \;.
\end{equation}

and the right hand side of the inequality given by \eqref{eq:choi_sym_thm} applied to the unital qubit Choi matrix \eqref{eq:unital_choi} is,

\begin{equation}
\begin{aligned}
1+\lambda_1^2+\lambda_2^2+\lambda_3^2\\ -\sqrt{(\lambda_1+\lambda_2+\lambda_3+1)(\lambda_1-\lambda_2-\lambda_3+1)(\lambda_1+\lambda_2-\lambda_3-1)(\lambda_1-\lambda_2+\lambda_3-1)}.
\end{aligned}
\end{equation}

We then apply a change of basis to \eqref{eq:unital_choi} to obtain \eqref{eq:bell_unital_choi} the Choi matrix for a unital qubit channel in the Bell basis,

\begin{equation}
\Choi_\mathbb{U}=\frac{1}{2}\begin{pmatrix}
\mu_0 & 0 & 0 & 0 \\
0 & \mu_1 & 0 & 0 \\
0 & 0 & \mu_2 & 0 \\
0 & 0 & 0 & \mu_3
\end{pmatrix}
\end{equation}

where, $\mu_0=(1+\lambda_1+\lambda_2+\lambda_3)$, $\mu_1=(1+\lambda_1-\lambda_2-\lambda_3)$,  $\mu_2=(1-\lambda_1+\lambda_2-\lambda_3)$, $\mu_3=(1-\lambda_1-\lambda_2+\lambda_3) $ as in  \eqref{eq:bell_choi}, the result follows. $_\blacksquare$\\

Because the Bell states diagonalize the unital qubit Choi matrix, we say that the matrix is \emph{Bell-diagonal}. The region of separability for the Bell diagonal states happens to be precisely when each $\mu_i \in [0, \frac{1}{2}],\; i=0,1,2,3$ \cite{cerf1998quantum,lang2010quantum,horodecki1996separability}. One can check that this affirms the fact that the set of entanglement breaking channels are a subset of the antidegradable channels. We note that the unital channel case has been previously characterized by other methods in \cite{cubitt2008structure}. It turns out that in the case of qubit channels, every unital channel is mixed unitary \cite{landau1993birkhoff}.

\subsection{Conditions for Antidegradable Qubit Channels}

The characterization of degradable qubit channels has been previously characterized in \cite{cubitt2008structure}. In the following let $\Z$ denote the auxiliary space in the minimal Stinespring representation for a channel $\Phi \in C(\X,\Y)$. Where both $\X$ and $\Y$ are complex Euclidean spaces of dimension two.

When $dim(\Z)=1$, the minimal Stinespring representation is represented by the isometry from the input space to the tensor product of the output with a one dimensional auxiliary space.

\begin{proposition}
If the auxiliary space $\Z$ in the minimal Stinespring representation is one dimensional, then any channel $\Phi \in C(\X,\Y)$ is degradable.
\end{proposition}

\textbf{Proof:}
Since $dim(\Z)=1$, implies that $rank(\mathcal{C}_\Phi)=1$, we can find Kraus representation with only one operator $A\in L(\X,\Y)$. It follows that $\Phi$ must be a unitary channel, since $A^*A=I$ by the trace preserving properties of every quantum channel. It is clear that the complementary channel of a unitary channel is the trace map, therefore $\Phi$ is degradable by the unitary invariance of the trace. $_\blacksquare$\\

If $dim(\Z)=2$, then $rank(\mathcal{C}_\Phi)=2$ and we have the that the output space and the environment space have equal dimension. This case is described by a theorem of Wolf and Perez which states that every qubit channel in the case of $rank(\mathcal{C}_\Phi)=2$, is either degradable or antidegradable \cite{wolf2007quantum}.

\begin{proposition}[Wolf \& Perez \cite{wolf2007quantum}]\label{2kraus}
Every qubit channel with Choi rank two is unitarily equivalent to the channel described by the following two Kraus operators,
\begin{equation}\label{eq:2kraus}
A_1=\begin{pmatrix}
\cos(\alpha) & 0 \\
0 & \cos(\beta)
\end{pmatrix}
\quad A_2=\begin{pmatrix}
0 & \sin(\beta) \\
\sin(\alpha) & 0
\end{pmatrix}\;.
\end{equation}

\end{proposition}

\begin{proposition}\label{rank2_anti}
Any rank 2 qubit channel with Kraus operators of the form \eqref{eq:2kraus}, is antidegradable if and only if $\cos(2\beta)\cos(2\alpha)\leq0$.
\end{proposition}

\textbf{Proof:}
We first remark that because of \eqref{eq:2kraus} the channel maps the identity matrix to the matrix,
\begin{equation}
\Lambda_2(I)=\begin{pmatrix}
 \cos^2(\alpha)+\sin^2(\beta) & 0 \\
0 & \sin^2(\alpha)+\cos^2(\beta)
\end{pmatrix}\;.
\end{equation}
The left hand side of \eqref{eq:choi_sym_thm} is then,
\begin{equation}
\begin{aligned}
\tr(\Lambda_2(I)^2)&=\left(\cos^2(\alpha)+\sin^2(\beta)\right)^2+\left(\sin^2(\alpha)+\cos^2(\beta)\right)^2\\
&=(\cos^2(\alpha)+1-\cos^2(\beta))^2+(1-\cos^2(\alpha)+\cos^2(\beta))^2\;,
\end{aligned}
\end{equation}

for the right hand side of \eqref{eq:choi_sym_thm} we can obtain the Choi matrix by,
\begin{equation}
 \Choi_{\Lambda_2}=vec(A_1)vec(A_1)^*+vec(A_2)vec(A_2)^*\;,
\end{equation}

\begin{equation}
\Choi_{\Lambda_2}=\begin{pmatrix}
\cos(\alpha)^2 & 0 & 0 & \cos(\alpha)\cos(\beta) \\
0 & \sin(\alpha)^2 & \sin(\alpha)\sin(\beta) & 0 \\
0 & \sin(\alpha)\sin(\beta) & \sin(\beta)^2 & 0 \\
\cos(\alpha)\cos(\beta) & 0 & 0 & \cos(\beta)^2\\
\end{pmatrix}\;.
\end{equation}

Remark, because the Choi matrix is not full rank, the determinant is zero, and we are only left with the term,
\begin{equation} 
\tr(\Choi_{\Lambda_2}^2)=2\cos^4(\alpha)+2\cos^4(\beta)+4(\cos^2(\alpha)-1)(\cos^2(\beta)-1)\;.
\end{equation}

To obtain an inequality with zero, we rewrite the antidegradability condition \eqref{eq:choi_sym_thm} as,
\begin{equation}
\begin{aligned}
&\tr(\Choi_{\Lambda_2}^2)-\tr(\Lambda_2(I)^2)\\
=&2\cos^4(\alpha)+2\cos^4(\beta)+4(\cos^2(\alpha)-1)(\cos^2(\beta)-1)\\
&-(1+\cos^2(\alpha)-\cos^2(\beta))^2-(1-\cos^2(\alpha)+\cos^2(\beta))^2\\
=&8\cos^2(\alpha)\cos^2(\beta)-4\cos^2(\alpha)-4\cos^2(\beta)+2\\
=&2\cos(2\beta)\cos(2\alpha)\leq0 \;,
\end{aligned}
\end{equation}
which is equivalent to,
\begin{equation}
\cos(2\beta)\cos(2\alpha)\leq0 \;. \;_\blacksquare
\end{equation}

\begin{proposition}\label{rank2_deg}
Any rank 2 qubit channel with Kraus operators of the form \eqref{eq:2kraus}, is degradable if and only if $\cos(2\beta)\cos(2\alpha)\geq0$
\end{proposition}

\textbf{Proof:}
This follows from the fact that when the rank of the Choi operator is two, any single qubit channel is either degradable or antidegradable \cite{wolf2007quantum}. $_\blacksquare$\\

Previously, in this case of rank 2 channels it was noted that a channel is antidegradable if $|\sin(\alpha+\beta)|\geq|\cos(\beta-\alpha)|$ and degradable if $|\sin(\alpha+\beta)|\leq|\cos(\beta-\alpha)|$ \cite{cubitt2008structure}.

\begin{proposition}\label{equivalence}
For rank 2 qubit channels the condition for degradibility $\cos(2\alpha)\cos(2\beta) \geq 0$, is equivalent to the condition $|\cos(\beta-\alpha)| \geq |\sin(\alpha+\beta)|$. Furthermore, the condition for antidegradability $\cos(2\alpha)\cos(2\beta) \leq 0$, is equivalent to the condition $|\cos(\beta-\alpha)| \leq |\sin(\alpha+\beta)|$.
\end{proposition}

\textbf{Proof:}
The first part follows straight from the calculation,
\begin{equation}
\begin{aligned}
\cos(2\alpha)\cos(2\beta)
&=\cos(2\alpha)(\cos^2(\beta)-\sin^2(\beta))\\
&=(\cos^2(\alpha)-\sin^2(\alpha))(\cos^2(\beta)-\sin^2(\beta))\\
&=\cos^2(\alpha)\cos^2(\beta)-\cos^2(\alpha)\sin^2(\beta)-\sin^2(\alpha)\cos^2(\beta)+\sin^2(\alpha)\sin^2(\beta)\\
&=\left(\cos(\alpha)\cos(\beta)+\sin(\alpha)\sin(\beta)\right)^2-\left(\sin(\alpha)\cos(\beta)+\cos(\alpha)\sin(\beta)\right)^2 \\
&=\cos^2(\beta-\alpha)-\sin^2(\alpha+\beta) \geq 0\\
&\Rightarrow \cos^2(\beta-\alpha)\geq \sin^2(\alpha+\beta) \;.
\end{aligned}
\end{equation}
The second statement follows by similar manipulation. $\blacksquare$\\

We remark that propositions \ref{rank2_anti} and \ref{rank2_deg} give the expressions for antidegradability and degradability are in terms of the parameters $\alpha$ and $\beta$. This allows for a simpler picture of degradability and antidegradability in this case, unlike the previous characterizations in \cite{cubitt2008structure} where the expressions are in terms of different combinations of $\alpha$ and $\beta$. The expression $\cos(2\alpha)\cos(2\beta)$ generates a \emph{checkerboard} like surface of peaks and valleys in 3 dimensions. Any point on the surface below the plane $\cos(2\alpha)\cos(2\beta)=0$ is an antidegradable channel while any point on the surface above the zero plane is a degradable channel.

If $dim(\Z)=3$, then $rank(\mathcal{C}_\Phi)=3$, again we know that the determinant will vanish from the left hand side of \eqref{eq:choi_sym_thm}. For the following let $\lambda=(\lambda_1,\lambda_2,\lambda_3)$, and $t=(t_1,t_2,t_3)$.

\begin{proposition}
A single-qubit channel $\Lambda_3$ with Choi rank 3 is antidegradable if its Choi operator is positive semidefinite and,
\begin{equation}
\frac{1}{\|\lambda\|+\|t\|}\geq \|\lambda\|-\|t\| \;,
\end{equation}
where $t$ and $\lambda$ are parameters of the map \eqref{eq:qubit_matrix}.
\end{proposition}

\textbf{Proof:}
We recall that we can obtain the matrix $\Lambda_3(I)$ from by determining the action of the channel on the matrix basis,

\begin{equation}
\Lambda_3(I)=\begin{pmatrix}
1+t_3 & t_1-it_2 \\
t_1+it_2 & 1-t_3 \\
\end{pmatrix}
\end{equation}

Applying \eqref{eq:qubit_choi} and noting that $\tr(\Lambda_3(I)^2)$ and the right hand side of \eqref{eq:choi_sym_thm} $\tr({\Choi_{\Lambda_3}}^2)$ reduce to the following

\begin{equation}\label{eq:r3_cond}
\begin{aligned}
2(1+t_1^2+t_2^2+t_3^2) &\geq \lambda_1^2+\lambda_2^2+\lambda_3^2+(t_1^2+t_2^2+t_3^2+1)\\
1+t_1^2+t_2^2+t_3^2 &\geq \lambda_1^2+\lambda_2^2+\lambda_3^2\\
1 &\geq \|\lambda\|^2-\|t\|^2\\
1&\geq (\|\lambda\|+\|t\|)(\|\lambda\|-\|t\|)\;. \quad _\blacksquare
\end{aligned}
\end{equation}

It is interesting that the right hand side of \eqref{eq:r3_cond} is 0 when $\|t\|=\|\lambda\|$.
\begin{corollary}
A unital single-qubit channel $\mathbb{U}_3$ with Choi rank 3 is antidegradable if and only if,
\begin{equation}
\|\lambda\| \leq 1\;,
\end{equation}
where $\lambda$ is the parameter of the map \eqref{eq:unital_qubit}.
\end{corollary}

\textbf{Proof:}
Recall a qubit channel is unital implies that $t=0$. $_\blacksquare$\\

Notice that one optimal condition among the unitary channels are points in the intersection between the ball $1=\lambda_1^2+\lambda_2^2+\lambda_3^2$ and the tetrahedron \eqref{eq:tetrahedron}.\\

When $dim(\Z)=4$, the Choi matrix for the qubit channels is full rank and the determinant is non-zero. The following inequality is obtained from the Choi matrix from the parametrized representation from \eqref{eq:qubit_choi}.

\begin{proposition}
A single qubit channel is antidegradable if its Choi operator is positive semidefinite and
\begin{equation}
\begin{aligned}
\sum_i \left(  \lambda_i^2(\lambda_i^2-2t_i^2-2)-t_i^2(t_i^2-2)\right)+2\sum_{i\neq j}\left(\lambda_i^2(\lambda_j^2-t_j^2)+(t_it_j)^2\right)-8\prod_i \left(\lambda_i\right)+1\\
\geq \left(\sum_i \left(\lambda_i^2-t_i^2\right)-1\right)^2
\end{aligned}
\end{equation}
holds, where $t$ and $\lambda$ are parameters of the map \eqref{eq:qubit_matrix}, and $1\leq i \leq 3$.
\end{proposition}

\textbf{Proof:}
This follows directly from our result \eqref{eq:choi_sym_thm} and direct calculations using the values from \eqref{eq:qubit_matrix} and \eqref{eq:qubit_choi} to obtain,

\begin{equation}\label{eq:dim4_anti}
\begin{aligned}
1+t_1^2+t_2^2+t_3^2 \geq \lambda_1^2+\lambda_2^2+\lambda_3^2\\
-\sqrt{\sum_i \big[  \lambda_i^2(\lambda_i^2-2t_i^2-2)-t_i^2(t_i^2-2)\big]+2\sum_{i\neq j}\big[\lambda_i^2(\lambda_j^2-t_j^2)+(t_it_j)^2\big]-8\prod_i\lambda_i+1}\;,
\end{aligned}
\end{equation}

We note that a better way to present this inequality is to move the negative square root term to the left hand side and group the other terms on the right hand side, after squaring both sides we obtain,

\begin{equation}\label{eq:anti_condition}
\begin{aligned}
LHS&=\sum_i \big[  \lambda_i^2(\lambda_i^2-2t_i^2-2)-t_i^2(t_i^2-2)\big]+2\sum_{i\neq j}\big[\lambda_i^2(\lambda_j^2-t_j^2)+(t_it_j)^2\big]-8\prod_i\lambda_i+1\\
RHS&= \left(\sum_i \left(\lambda_i^2-t_i^2\right)-1\right)^2
\end{aligned}
\end{equation}

for which we have an antidegradable channel whenever $LHS\geq RHS$, hence our result follows. $_\blacksquare$\\

\section{Applications to Qubit Quantum Channels}

In this section we will discuss a few examples of qubit channels and their degradability or antidegradable criteria.

\subsection{Dephasing Channel}

For single-qubit channels degradable channels only exist if the rank of the Choi operator is less than or equal to two \cite{cubitt2008structure}. Since the dephasing maps are a subset of the degradable channels it means we require at most two Kraus operators to describe any single-qubit dephasing channel.

\begin{remark}[Wolf \& Perez \cite{wolf2007quantum}]
The rank 2 qubit channel \eqref{eq:2kraus} is a dephasing channel when $\alpha=\beta$,
\begin{equation}
B_1=\begin{pmatrix}
\cos(\alpha) & 0 \\
0 & \cos(\alpha)
\end{pmatrix}
\quad B_2=\begin{pmatrix}
0 & \sin(\alpha) \\
\sin(\alpha) & 0
\end{pmatrix}\;.
\end{equation}
\end{remark}

\textbf{Proof:} When $\alpha=\beta$ this channel will be dephasing by the fact that its Kraus operators are simultaneously diagonalizable. Because $A_2$ is hermitian we can diagonalize it with some change of basis which will not affect $A_1$, because $A_1$ is a multiple of the identity matrix, hence both operators are diagonal in the same basis. $_\blacksquare$\\

We remark that degradability of this channel easily follows from our results, because the dephasing condition $\alpha=\beta$, implies that proposition \ref{rank2_deg} is always satisfied.

\subsection{Amplitude Damping Channel}

\begin{remark}[Wolf \& Perez \cite{wolf2007quantum}]
The rank 2 qubit channel \eqref{eq:2kraus} is the amplitude damping channel when $\beta=0$, is given by the following two Kraus operators,

\begin{equation}
C_1=\begin{pmatrix}
\cos(\alpha) & 0 \\
0 & 1
\end{pmatrix}
\quad C_2=\begin{pmatrix}
0 & 0 \\
\sin(\alpha) & 0
\end{pmatrix}\;.
\end{equation}
\end{remark}

\begin{proposition}
The qubit amplitude damping channel is antidegradable for $\cos(2\alpha)\leq0$, and degradable for $\cos(2\alpha)\geq 0$.
\end{proposition}

\textbf{Proof:} This follows from the condition for amplitude damping channels of $\beta=0$ and propositions \ref{rank2_deg} and \ref{rank2_anti}. $_\blacksquare$\\

\subsection{Depolarizing Channel}

Another important example is the depolarizing channel. The depolarizing channel act as the identity channel on a density operator with some probability $(1-p)$ and act as the \emph{completely depolarizing} channel with the probability $p$.

\begin{definition}
Let $\rho$ be a density operator the completely depolarizing channel is defined as,
\begin{equation}\label{eq:compdepo}
\Omega(\rho)=\frac{I}{d}\tr(\rho)\;,
\end{equation}
where $d$ is the dimension of $\rho$.
\end{definition}

\begin{definition}
Let $\rho$ be a density operator and $p \in (0,1)$ the depolarizing channel is defined as,
\begin{equation}\label{eq:depo}
\Upsilon(\rho)=(1-p)\rho+p\frac{I}{d}\tr(\rho)\;,
\end{equation}
where $d$ is the dimension of $\rho$.
\end{definition}

\begin{proposition}
The depolarizing channel is antidegradable for $p\geq \tfrac{1}{3}$.
\end{proposition}

\textbf{Proof:} Because the Choi correspondence is a linear vector space isomorphism we can find the Choi matrix of the depolarizing channel in terms of $p$, simply by taking the convex combination of the identity and completely depolarizing Choi representations,

\begin{equation}\label{eq:dep_choi}
\Choi_\Upsilon=(1-p)\Choi_{\mathcal{I}}+p\,\Choi_\Omega=\begin{pmatrix}
1-p & 0 & 0 & 1-p\\
0 & 0 & 0 & 0\\
0 & 0 & 0 & 0\\
1-p & 0 & 0 & 1-p
\end{pmatrix}+\begin{pmatrix}
\frac{p}{2} & 0 & 0 & 0\\
0 & \frac{p}{2} & 0 & 0\\
0 & 0 & \frac{p}{2} & 0\\
0 & 0 & 0 & \frac{p}{2}
\end{pmatrix}=\begin{pmatrix}
1-\frac{p}{2} & 0 & 0 & 1-p\\
0 & \frac{p}{2} & 0 & 0\\
0 & 0 & \frac{p}{2} & 0\\
1-p & 0 & 0 & 1-\frac{p}{2}
\end{pmatrix}\;,
\end{equation}
where $p \in [0,1]$.

We note that this channel is unital and hence the left hand side of our inequality \eqref{eq:choi_sym_thm} is 2. We note that the depolarizing channel Choi matrix \eqref{eq:dep_choi} is of the form of a unital channel, see \eqref{eq:unital_choi}. Hence, the qubit depolarizing channel is antidegradable whenever,

\begin{equation}
2 \geq 3p^2 -  6p + 4- \sqrt{-p^3(3p - 4)} \;,
\end{equation}
which on the interval $[0,1]$ is when,
\begin{equation}
 p \geq \frac{1}{3}  \;.\;_\blacksquare \\
\end{equation}

Furthermore, it was shown in \cite{holevo2012book} that the depolarizing channel is entanglement breaking for $p \geq \tfrac{2}{3}$.

\begin{remark}
The qubit depolarizing channel is antidegradable but not entanglement breaking for $p \in [\tfrac{1}{3},\tfrac{2}{3})$.\\
\end{remark}

This reaffirms the results in \cite{Leung:2015aa} that the complementary channel of the qubit depolarizing channel has positive capacity for values of $p$, in addition to being degradable. One may hope to find optimal codes for the depolarizing channel, in the sense of finding codes where $Q(\Upsilon_2)>0$ for $p= \tfrac{1}{3}-\epsilon$.

\section{Self-Complementary Qubit Channels}

It is known that degradable qubit channels only exist when $dim(\Z)\leq 2$ \cite{cubitt2008structure}. In addition, antidegradable qubit channels only exist for $dim(\Z)\geq 2$ \cite{cubitt2008structure}. We now investigate the conditions for a qubit channel which is both degradable and antidegradable.

\begin{proposition}
A qubit channel is degradable and antidegradable if $|\sin(\alpha+\beta)|=|\cos(\beta-\alpha)|$. Where the channel has the Kraus operators
\begin{equation}\label{eq:kraus_rank2}
K_1=\begin{pmatrix}
\cos(\alpha) & 0 \\
0 & \cos(\beta)
\end{pmatrix}
\quad K_2=\begin{pmatrix}
0 & \sin(\beta) \\
\sin(\alpha) & 0
\end{pmatrix}\;.
\end{equation}
\end{proposition}

\textbf{Proof:} When the Choi rank is 2 every channel is either degradable and antidegradable, then by the form of our conditions for degradability and antidegradability (propositions \ref{rank2_deg} and \ref{rank2_anti}) the result follows. $_\blacksquare$\\

\begin{proposition}
A qubit channel with $dim(\Z)=2$, is self complementary if $|\cos(\beta)|=|\sin(\beta)|$.
\end{proposition}

\textbf{Proof}:
Recalling when $dim(\Z)=2$, by proposition \ref{2kraus} we can write the channel in terms of the two Kraus operators \eqref{eq:kraus_rank2}. We remark that the Kraus operators ensure that our map will be completely positive for appropriate regions of $\alpha$ and $\beta$. We can then construct the partial isometry of the channels Stinespring representation,
\begin{equation}
V=K_1\otimes e_1+K_2\otimes e_2=\begin{pmatrix}
\cos(\alpha) & 0 \\
0 & \sin(\beta)\\
0 & \cos(\beta) \\
\sin(\alpha) & 0
\end{pmatrix}\;.
\end{equation}

We remark that a channel with the Stinespring isometry $V \in \mathcal{V}(\X,\Y \otimes \Z)$, is self-complementary if
\begin{equation}
\tr_\Z(V\rho V^*)=\tr_\Y(V\rho V^*)\;,
\end{equation}
for all $\rho \in \mathcal{D}(\X)$.\\

Now, let $\rho=\begin{pmatrix}
x & z \\
\bar{z} & y
\end{pmatrix} \in \mathcal{D}(\X)$. Now remark the action of the isometry $M=V\rho V^*$,

\begin{equation}
\begin{aligned}
M=\begin{pmatrix}
\cos(\alpha) & 0 \\
0 & \sin(\beta)\\
0 & \cos(\beta) \\
\sin(\alpha) & 0
\end{pmatrix}
\begin{pmatrix}
x & z \\
\bar{z} & y
\end{pmatrix}
\begin{pmatrix}
\cos(\alpha) & 0 & 0 & \sin(\alpha) \\
0 & \sin(\beta) & \cos(\beta) & 0\\
\end{pmatrix}\\
M=\begin{pmatrix}
x\cos(\alpha)^2 & z\cos(\alpha)\sin(\beta) & z\cos(\alpha)\cos(\beta) & x\cos(\alpha)\sin(\alpha)\\
\bar{z}\cos(\alpha)\sin(\beta) & y\sin(\beta)^2 & y\cos(\beta)\sin(\beta) & \bar{z}\sin(\beta)\sin(\beta)\\
\bar{z}\cos(\alpha)\cos(\beta) & y\cos(\beta)\sin(\beta) & y\cos(\beta)^2 & \bar{z}\sin(\alpha)\cos(\beta)\\
 x\cos(\alpha)\sin(\alpha) & z\sin(\beta)\sin(\alpha) & z\cos(\alpha)\cos(\beta) & x\sin(\alpha)^2\\
\end{pmatrix}\;.
\end{aligned}
\end{equation}

Let us label the above matrix $M$ as four $2 \times 2$ sub-matrices $
\begin{pmatrix} M_1 & M_2 \\ M_3 & M_4\\ \end{pmatrix}\;.$

The action of tracing out partial systems $\Y$ and $\Z$ amounts to the following block matrix calculations
\begin{equation}
\begin{aligned}
\tr_\Y(M)&=(M_1+M_4)\\&=\begin{pmatrix}
x\cos(\alpha)^2+ y\cos(\beta)^2 & z\cos(\alpha)\sin(\beta)+\bar{z}\sin(\alpha)\cos(\beta)\\
\bar{z}\cos(\alpha)\sin(\beta)+z\sin(\alpha)\cos(\beta) & y\sin(\beta)^2+x\sin(\alpha)^2\\
\end{pmatrix}\;,
\end{aligned}
\end{equation}
and
\begin{equation}
\begin{aligned}
\tr_\Z(M)&=\begin{pmatrix} \tr(M_1) & \tr(M_2)\\ \tr(M_3) & \tr(M_4)\\ \end{pmatrix}\\&=\begin{pmatrix}
x\cos(\alpha)^2+y\sin(\beta)^2 & z\cos(\alpha)\cos(\beta)+ \bar{z}\sin(\alpha)\sin(\beta)\\
\bar{z}\cos(\alpha)\cos(\beta)+ z\sin(\alpha)\sin(\beta)&  y\cos(\beta)^2+x\sin(\alpha)^2 &\\
\end{pmatrix}\;.
\end{aligned}
\end{equation}

Notice that if $\sin(\beta)=\cos(\beta)$ the matrix $\tr_\Z(M)$ is the same as $\tr_\Y(M)$. $_\blacksquare$\\

Remark that any self-complementary channel is both degradable and antidegradable, because if $\Phi=\widetilde{\Phi}$ then the identity map degrades both the channel and its complementary channel. We note that there are two solutions to the self complementary condition $|\sin(\alpha+\beta)|=|\cos(\beta-\alpha)|$, first $|\cos(\beta)|=|\sin(\beta)|$ and secondly $|\sin(\alpha)|=|\cos(\beta)|$. We remark that the first solution yields the set of self complementary channels and the second solution yields a the set of degradable and antidegradable channels which are not self-complementary.

\section{Conclusion}

The application of this research could be used to find good codes for noisy qubit channels. Degradable channels were used in \cite{Leditzky:2017aa} to derive a formula for the channel capacity in a low-noise regime to leading orders, perhaps these methods may be helpful in those directions. Finding qubit channels which are antidegradable but not entanglement breaking may provide a framework for finding channels in which shared entanglement generation may be possible even when these channels alone have zero coherent information. Lastly, it is not apparent that the set of Choi matrices which satisfy \eqref{eq:anti_cor} is convex, without relying on the convexity of the set of symmetrically extendible operators \cite{myhr2009spectrum}. It would be interesting to characterize the extreme points of this set and determine the properties of the resulting qubit channels, as they may be useful in studying the coherent information. Mathematically, the part of the inequality which provides difficulty in determining convexity, is the function $\sqrt{\det({\Choi_\Phi})}$, where $\Choi_\Phi$ is a self-adjoint $4 \times 4$ positive semi-definite matrix. In general the function $\det(A)^{\frac{1}{2}}$ of a $4 \times 4$ matrix $A$ is not operator concave. We recall that if $A$ is an $n \times n$ matrix then $\det(A)^{\frac{1}{n}}$ is operator concave \cite{bhatia1997book}, however, there does not appear to be a way to extend this result to show operator concavity of the function in this case \cite{olkin2016inequalities}.

\section{Acknowledgements}

This research was conducted as part of the first authors M.Sc while attending the University of Guelph and the Institute for Quantum Computing, he would like to thank Rajesh Pereira, Bei Zeng, Nengkun Yu, and Joel Klassen for many helpful discussion.

\bibliographystyle{plain}

\begin{thebibliography}{10}

\bibitem{bhatia1997book}
R.~Bhatia.
\newblock {\em Matrix Analysis}.
\newblock Springer, 1997.

\bibitem{braun2014universal}
Daniel Braun, Olivier Giraud, Ion Nechita, et~al.
\newblock A universal set of qubit quantum channels.
\newblock {\em Journal of Physics A: Mathematical and Theoretical},
  47(13):135302, 2014.

\bibitem{cerf1998quantum}
Nicolas~J. Cerf.
\newblock Quantum cloning and the capacity of the {Pauli} channel.
\newblock {\em arXiv preprint quant-ph/9803058}, 1998.

\bibitem{chen2014symmetric}
Jianxin Chen, Zhengfeng Ji, David Kribs, Norbert L{\"u}tkenhaus, and Bei Zeng.
\newblock Symmetric extension of two-qubit states.
\newblock {\em Physical Review A}, 90(3):032318, 2014.

\bibitem{choi1975completely}
Man-Duen Choi.
\newblock Completely positive linear maps on complex matrices.
\newblock {\em Linear algebra and its applications}, 10(3):285--290, 1975.

\bibitem{cubitt2008structure}
Toby~S. Cubitt, Mary~Beth Ruskai, and Graeme Smith.
\newblock The structure of degradable quantum channels.
\newblock {\em Journal of Mathematical Physics}, 49(10):102104, 2008.

\bibitem{devetak2005capacity}
Igor Devetak and Peter~W. Shor.
\newblock The capacity of a quantum channel for simultaneous transmission of
  classical and quantum information.
\newblock {\em Communications in Mathematical Physics}, 256(2):287--303, 2005.

\bibitem{fannes1988symmetric}
Mark Fannes, J.T. Lewis, and Andr{\'e} Verbeure.
\newblock Symmetric states of composite systems.
\newblock {\em Letters in mathematical physics}, 15(3):255--260, 1988.

\bibitem{fujiwara1999one}
Akio Fujiwara and Paul Algoet.
\newblock One-to-one parametrization of quantum channels.
\newblock {\em Physical Review A}, 59(5):3290, 1999.

\bibitem{holevo2007complementary}
Alexander~Semenovich Holevo.
\newblock Complementary channels and the additivity problem.
\newblock {\em Theory of Probability \&amp; Its Applications}, 51(1):92--100,
  2007.

\bibitem{holevo2012book}
Alexander~Semenovich Holevo.
\newblock {\em Quantum Systems, Channels, Information}.
\newblock Berlin: De Gruyter, 2012.

\bibitem{horn1991topics}
R.~Horn and C.~Johnson.
\newblock {\em Topics in Matrix Analysis}.
\newblock Cambridge university press, 1991.

\bibitem{horodecki2003entanglement}
Michael Horodecki, Peter~W. Shor, and Mary~Beth Ruskai.
\newblock Entanglement breaking channels.
\newblock {\em Reviews in Mathematical Physics}, 15(06):629--641, 2003.

\bibitem{horodecki1996separability}
Micha{\l} Horodecki, Pawe{\l} Horodecki, and Ryszard Horodecki.
\newblock Separability of mixed states: necessary and sufficient conditions.
\newblock {\em Physics Letters A}, 223(1):1--8, 1996.

\bibitem{jamiolkowski1972linear}
Andrzej Jamio{\l}kowski.
\newblock Linear transformations which preserve trace and positive
  semidefiniteness of operators.
\newblock {\em Reports on Mathematical Physics}, 3(4):275--278, 1972.

\bibitem{king2005properties}
Christopher King, Keiji Matsumoto, Michael Nathanson, and Mary~Beth Ruskai.
\newblock Properties of conjugate channels with applications to additivity and
  multiplicativity.
\newblock {\em arXiv preprint quant-ph/0509126}, 2005.

\bibitem{king2001minimal}
Christopher King and Mary~Beth Ruskai.
\newblock Minimal entropy of states emerging from noisy quantum channels.
\newblock {\em IEEE Transactions on information theory}, 47(1):192--209, 2001.

\bibitem{landau1993birkhoff}
L.J. Landau and R.F. Streater.
\newblock On {Birkhoff's} theorem for doubly stochastic completely positive
  maps of matrix algebras.
\newblock {\em Linear algebra and its applications}, 193:107--127, 1993.

\bibitem{lang2010quantum}
Matthias~D. Lang and Carlton~M. Caves.
\newblock Quantum discord and the geometry of {Bell}-diagonal states.
\newblock {\em Physical review letters}, 105(15):150501, 2010.

\bibitem{Leditzky:2017aa}
Felix Leditzky, Debbie Leung, and Graeme Smith.
\newblock Quantum and private capacities of low-noise channels.
\newblock {\em arXiv preprint}, arXiv:1705.04335, 2017.

\bibitem{Leung:2015aa}
Debbie Leung and John Watrous.
\newblock On the complementary quantum capacity of the depolarizing channel.
\newblock {\em arXiv preprint}, arXiv:1705.04335, 2015.

\bibitem{myhr2009spectrum}
Geir~Ove Myhr and Norbert L{\"u}tkenhaus.
\newblock Spectrum conditions for symmetric extendible states.
\newblock {\em Physical Review A}, 79(6):062307, 2009.

\bibitem{olkin2016inequalities}
Ingram Olkin and Albert~W. Marshall.
\newblock {\em Inequalities: theory of majorization and its applications},
  volume 143.
\newblock Academic press, 2016.

\bibitem{ruskai2002analysis}
Mary~Beth Ruskai, Stanislaw Szarek, and Elisabeth Werner.
\newblock An analysis of completely-positive trace-preserving maps on {M}2.
\newblock {\em Linear Algebra and its Applications}, 347(1):159--187, 2002.

\bibitem{stinespring1955positive}
W.~Forrest Stinespring.
\newblock Positive functions on {C*}-algebras.
\newblock {\em Proceedings of the American Mathematical Society},
  6(2):211--216, 1955.

\bibitem{werner1989application}
Reinhard~F. Werner.
\newblock An application of {Bell's} inequalities to a quantum state extension
  problem.
\newblock {\em Letters in Mathematical Physics}, 17(4):359--363, 1989.

\bibitem{wolf2007quantum}
Michael~M. Wolf and David Perez-Garcia.
\newblock Quantum capacities of channels with small environment.
\newblock {\em Physical Review A}, 75(1):012303, 2007.

\end{thebibliography}

\end{document}